\documentclass[pss]{wiley2sp}
\usepackage{w-greek}
\usepackage{amsmath}% for pmb
\usepackage{bm}

\begin{document}

% If \titlefigure is given, it takes as its mandatory parameter the
% name (without extension) of some figure file. The file must be at a
% place where LaTeX will find it; in most cases it will be best to put
% it right where the respective TeX file resides.
%
% The optional parameter can be used to pass additional parameters
% (key-value pairs) to the underlying figure inclusion mechanism like
% height (see below), width (which defaults to local \linewidth and
% should always be given in some decimal fraction of this, if needed),
% rotations, bounding box manipulations and others; see the 'LaTeX
% standard graphics and color packages documentation' for this, to be
% found at
% <http://tug.ctan.org/tex-archive/macros/latex/required/graphics/grfguide.pdf>.
%
% Accepted figure file formats depend on which LaTeX flavour is used.
% Classic LaTeX is always able to use Encapsulted Postscript (EPS);
% PDFLaTeX can't use this but accepts PDF, JPG, PNG, and GIF formats.
% Certain TeX implementations such as MikTeX support additional
% formats, but keep in mind that usage of this feature will sacrifice
% portability.
%

\title{Ultralong-range radiative excitation transfer between quantum dots in a planar microcavity}

\titlerunning{Ultralong-range excitation transfer between QDs in a microcavity }

\author{%
  Guillaume Tarel\textsuperscript{\textsf{\bfseries \Ast}},
  Gaetano Parascandolo,
  Vincenzo Savona}

\authorrunning{G. Tarel et al.}

%\def\mailname{Standard is "Corresponding author"}
%E-mail-address of corresponding author
\mail{e-mail
  \textsf{guillaume.tarel@epfl.ch}, Phone
  +41-21-6933423, Fax +41-21-6935419}

\institute{%
 Institut de Physique Th\'eorique, Ecole Polytechnique F\'ed\'erale de Lausanne (EPFL), CH-1015 Lausanne, Switzerland}

\received{XXXX, revised XXXX, accepted XXXX}
\published{XXXX}

 %%PACS-Numbers

\abstract{We study the system of two quantum dots lying on the central plane of a planar semiconductor microcavity. By solving the full Maxwell problem, we demonstrate that the rate of resonant excitation transfer between the two dots decays as $d^{-1/2}$ as a function of the distance $d$ at long distance. This very long-range mechanism is due to the leaky and guided modes of the microcavity, which act as effective radiative transfer channels. At short distance, the $d^{-3}$ dependence of the F\"orster mechanism, induced by the electrostatic dipole-dipole interaction, is recovered.}

\maketitle

\section{Introduction}

Semiconductor quantum dots (QDs) are promising candidates for a solid-state implementation of quantum 
information technology. To this purpose, both orbital and spin electronic degrees of freedom might be exploited. 
One of the major challenges along this route consists in achieving a controlled quantum state preparation and 
transfer of quantum excitation between different subsystems. This latter task is particularly difficult if the 
subsystems are spatially separated. Very effective control schemes might be based on controlled coupling of QD 
electronic degrees of freedom with photons. Locally, strong coupling between an electron-hole pair state in a QD 
and a resonant mode of an optical cavity was recently demonstrated \cite{Reithmaier2004}, as the semiconductor 
counterpart of atom-cavity strong coupling in cavity quantum electrodynamics. Thanks to optical control and detection, 
strong coupling should allow full control of the QD quantum state. For QDs lying separate in space, 
the coupling to photons is the most effective way to achieve controlled excitation transfer, particularly by taking 
advantage of resonant photonic structures \cite{Reitzenstein2006,Englund2007,Hughes2007}. The same mechanism could be %equally
 used to transfer the electron spin state \cite{Quinteiro2006}.

In order to get better insight into these mechanisms, the fundamental nature of the photon-mediated transfer must be studied. At short range, electrostatic coupling between the QD dipole moments results in the F\"orster resonant energy transfer \cite{Govorov2005}, with a probability amplitude decaying as the third power of the QD distance. Recently, we have demonstrated \cite{Parascandolo2005,Parascandolo2007} that the radiative field produces a long-range mechanism dominating over the F\"orster transfer at long distance. The mechanism can be pictured as the emission of a photon by a QD and reabsorption by a distant QD. It decays as the inverse of the QD distance, as expected from spherical wave propagation, and could be exploited to engineer quantum correlations \cite{Hughes2006}. Experimental evidence of it was recently found \cite{Scheibner2007}.

In this work we extend the analysis of radiative excitation transfer to the case where QDs lie in a planar semiconductor microcavity (MC) with distributed Bragg reflectors (DBRs). In this configuration, the transfer takes place through both the main resonant mode of the MC and guided modes originating from the DBR structure. We present a semiclassical formalism in which the QD optical susceptibility is coupled to the full Maxwell equations. We prove that the probability amplitude of the spontaneous excitation transfer decays as $d^{-1/2}$ as a function of the distance between QDs $d$. Depending on the direction of the electron-hole dipole moment and on the QD-cavity detuning, this ultralong-range transfer can be enhanced or inhibited, thanks to the resonant character of the MC. This principle might be exploited to control the transfer mechanism, e.g. by means of laser-controlled ultrafast cavity tuning \cite{Fushman2007}.

\section{Theory}

We consider the system of two QDs embedded in a planar MC with DBRs. We assume a $\lambda$-MC with the QDs lying on 
its central plane, where the electric field has maximum amplitude. We assume the ${\bf z}$-axis as the MC 
vertical axis, and the ${\bf x}$-axis as the axis joining the two QDs. 
The method follows directly the one adopted in Ref. \cite{Parascandolo2005}. 
In the limit of low QD excitation, the excitation transfer is governed by linear optics and is described by the 
Maxwell equation for the electric field $\bm{\mathcal{E}}$ coupled to the linear optical susceptibility of the QDs
\begin{eqnarray} \label{Max}
&&\bm{\nabla}\wedge\bm{\nabla}\wedge
{\bm{\mathcal E}}\left({\bm{r}},\omega\right)-\frac{\omega^2}{c^2}
\bigg[\epsilon_\infty {\bm{\mathcal E}}\left({\bm{r}},\omega\right)
\\
&&\left.+4\pi\int d{\bm{r}}^\prime
\hat{\bm{\chi}}\left({\bm{r}},{\bm{r}}^\prime,\omega\right)
\cdot {\bm{\mathcal E}}\left({\bm{r}}^\prime,\omega\right)\right]=0
\,,\nonumber
\end{eqnarray}
where $\bm{r}$ is the 3-D position vector, $\epsilon_\infty$ the dielectric constant of the MC central layer, 
and $\hat{\bm{\chi}}$ the $3\times3$ linear optical susceptibility tensor of the QD subsystem. By assuming the strongest 
confinement of electrons and holes along the ${\bf z}$-direction of the QD, we can assume that the same selection rules for the 
interband transition as in a quantum well hold. Then, by restricting to the electron-heavy-hole transition in a crystal of 
cubic symmetry (e.g. InGaAs), the susceptibility tensor is expressed as
\begin{equation} \label{Chi}
\hat{\bm{\chi}}\left({\bf r},{\bf r}^\prime,\omega\right)=
\frac{\mu_{cv}^2}{\hbar}\sum_{\alpha=1,2}
\frac{\Psi^{ }_{\alpha}\left({\bf r},{\bf r}\right)
\Psi^*_{\alpha}\left({\bf r^\prime},{\bf r^\prime}\right)}
{\omega_{\alpha}-\omega-i0^+}
\left(
\begin{array}{ccc}
1 & 0 & 0 \\
0 & 1 & 0 \\
0 & 0 & 0
\end{array}
\right)\,,
\end{equation}
where $\mu_{cv}$ is the Bloch part of the interband dipole matrix element and $\Psi^{ }_{\alpha}({\bf r}_e,{\bf r}_h)$ is 
the electron-hole-pair envelope wave function in the $\alpha$-th QD. Given the strongly resonant character of the radiative 
transfer mechanism, we have safely restricted our analysis to the ground e-h pair state of each QD. For the QD wave function, 
we neglect Coulomb correlation and assume the factorized form:
%factorized electron and hole degrees of freedom, as well as factorized in-plane and $z$ motion
\begin{equation}
\Psi^{ }_{\alpha}({\bf r}_e,{\bf r}_h)=f_e(z_e)f_h(z_h)\phi_{e,\alpha}(\bm{\rho}_e)\phi_{h,\alpha}(\bm{\rho}_h)\,.
\label{wavef}
\end{equation}
For the calculations, the in-plane wave functions are modeled as isotropic Gauss functions with standard deviations $\sigma_{e}$ 
and $\sigma_{h}$. We further assume $\delta$-confinement along the ${\bf z}$-direction by taking $f_j^2(z)=\delta(z)$. 
This very simplified model can catch the essential features of the QD optical transition, relevant to the excitation transfer mechanism.

Under these assumptions, the steps leading to a coupled mode equation are formally the same as in Ref. \cite{Parascandolo2005}. Maxwell equation (\ref{Max}) ca be cast into an integral Dyson equation \cite{Martin1998} which finally reads
\begin{equation}
\label{projected_Dyson}
{\bf E}_\alpha(\omega)={\bf E}^0_\alpha+\sum_\beta
\frac{\hat{\bf G}_{\alpha\beta}(\omega)}
{\omega_\beta-\omega-i0^+}{\bf E}_\beta(\omega)\,,
\end{equation}
where $\hbar\omega_\alpha$ is the transition energy of the $\alpha$-th QD and
\begin{equation}
{\bf E}_\alpha(\omega)=\sum_{\bf k}{\psi^{ }_{\alpha,{\bf k}}}{\bf E}_k(\omega)\,,
\end{equation}
\begin{equation}\label{G_ab}
\hat{\bf G}_{\alpha\beta}(\omega)=4\pi\frac{k_0^2}{\epsilon_\infty}
\frac{\mu_{cv}^2}{\hbar}\sum_{\bf k}{\psi^{ }_{\alpha,{\bf k}}}
\hat{\bf G}_k(\omega){\psi^{*}_{\beta,{\bf k}}}\,.
\end{equation}
Here, $\psi_{\alpha,{\bf k}}$ is the two--dimensional Fourier transform 
of $\psi_\alpha({\bm{\rho}})=\phi_{e,\alpha}(\rho)\phi_{h,\alpha}(\rho)$, which 
is assumed to be centered at the position $\bm{R}_\alpha$ of the $\alpha$-th QD on the plane, and ${\bf E}_{\bf k}(\omega,z)$ 
is the Fourier-transform, with respect to the in-plane coordinate $\pmb{\rho}$, of the in-plane projection ${\bf E}(\omega,{\bf r})$ 
of the electric field $\bm{\mathcal{E}}(\omega,{\bf r})$.  ${\bf E}^0_\alpha$ is defined as ${\bf E}_\alpha$ and accounts for a possible external field ${\bf E}^0$. We have further defined $k_0=(\omega/c)\sqrt{\epsilon_\infty}$. Eqs. (\ref{projected_Dyson})-(\ref{G_ab}) are evaluated at $z=0$, as the $\delta$-assumption for the $z$-confinement allows to carry out the $z$-integral in (\ref{Max}) explicitly.

The central quantity of the present analysis is the in-plane Green's tensor of the photon $\hat{\bf G}_k(\omega,z)$, which is defined as the Green's tensor of the two coupled equations for ${\bf E}_{\bf k}(\omega,z)$, resulting from Eq. (\ref{Max}) after having eliminated the uncoupled ${\bf z}$-component of the electric field. Notice that this quantity depends on $k=|{\bf k}|$ only, because of the circular invariance of the Maxwell problem. For QDs in a bulk dielectric medium, $\hat{\bf G}_k(\omega,z)$ is given by a simple analytical expression \cite{Parascandolo2005}. In a MC, we have shown \cite{Savona1996} that an analytical expression still holds. The tensor is diagonal on the basis of transverse (TE) and longitudinal (TM) electric field polarization with respect to the direction of ${\bf k}$. In particular, at $z=0$ we have
\begin{eqnarray}
\hat{\bf G}_k(\omega)&=&\frac{i}{2k_0^2}
\left(\begin{array}{cc}
\frac{k_0^2}{k_z}&0\\
0&k_z
\end{array}\right)
\frac{\left(1+r^t_{k}e^{ik_zL_c}\right)\left(1+r^b_{k}e^{ik_zL_c}\right)}{\left(1-r^t_{k}r^b_{k}e^{2ik_zL_c}\right)}\,,
\label{Gcav}
\end{eqnarray}
where $k_z=\sqrt{k_0^2-k^2}$ and $r^j_{k}(\omega)$ ($j=t,b$) are the complex reflection coefficients of the top and bottom DBR respectively, that can be computed by means of a transfer-matrix technique \cite{Savona1995}. We define $g^L_k(\omega)$ and $g^T_k(\omega)$ as the diagonal elements of $\hat{\bf G}_k(\omega)$.
They represent the propagators for TM and TE modes respectively. In the limit $r^j_{k}=0$, the in-plane propagator in a homogeneous medium \cite{Parascandolo2005} is correctly recovered from Eq. (\ref{Gcav}). By replacing the (normalized) Gauss wave functions of the QD states and (\ref{Gcav}) into (\ref{G_ab}), straightforward algebra leads to a simple expression for the rates $\hat{G}_{\alpha\beta}(\omega)$. In particular, by assuming that the two QDs lie along the ${\bf x}$-axis, the interdot tensor $\hat{G}_{12}(\omega)$ can be expressed in the $(x,y)$-basis as (for simplicity we omit the $\omega$-dependence from the notation)
\begin{eqnarray}
\lefteqn{\hat{G}_{12}=\frac{8k_0^2\mu_{cv}^2}{\hbar\epsilon_\infty}\frac{\sigma^4}{\sigma_e^2\sigma_h^2}\int_0^\infty dk ~k~ e^{-k^2\sigma^2}
\left(\begin{array}{cc}
g^x_k&0\\
0&g^y_k
\end{array}\right)\,,\label{G12}}\\
&&g^x_k(d)=g^L_k\left(\frac{J_1(kd)}{kd}-J_2(kd)\right)+g^T_k\frac{J_1(kd)}{kd}\,,\label{gx}\\
&&g^y_k(d)=g^T_k\left(\frac{J_1(kd)}{kd}-J_2(kd)\right)+g^L_k\frac{J_1(kd)}{kd}\,,\label{gy}
\end{eqnarray}
where $d$ is the distance between the two QDs and $\sigma^2=\sigma_e^2\sigma_h^2/(\sigma_e^2+\sigma_h^2)$. Eq. (\ref{G12}) indicates that, in the case of two isotropic QDs, the ${\bf x}$- and ${\bf y}$-components of the e-h polarization vector are separately conserved. This is in general not true for an anisotropic QD, and the off-diagonal transfer rates will depend on the degree of anisotropy of the wave functions.

\begin{figure}[ht]
  \centerline{\includegraphics*[width=1.0\linewidth]{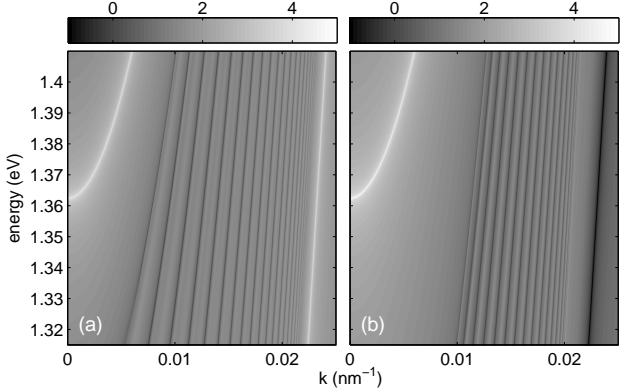}}
  \caption{(a): Logarithmic grayscale plot of $|g^T_k(\omega)|$. (b): Same for $|g^L_k(\omega)|$.}
  \label{fig2}
\end{figure}

\section{Results~and~discussion}

We study the 
%specific
case of a GaAs MC with top (air) and bottom (substrate) DBRs made of 20 and 19.5 pairs of GaAs/AlAs $\lambda/4$-layers respectively. 
We take $n_1=3$ and $n_2=3.5$ as the AlAs and GaAs refraction indices, and $L_c=260$ nm as the cavity layer thickness. We assume two 
identical QDs with $\sigma_e=\sigma_h=5$ nm and square dipole moment $\mu_{cv}^2=0.66~\mbox{meV}\cdot\mbox{nm}^{3}$, 
corresponding to a Kane energy of 21.5 eV as in InAs.

%which corresponds to a Kane energy of 21.5 eV as in InAs.

By inspection of Eqs. (\ref{G12})-(\ref{gy}), we can characterize the long-range behaviour of the radiative transfer mechanism. 
We first point out that Bessel functions $J_n(x)$ decay asymptotically as $x^{-1/2}$. Hence, the $J_2$ contributions in (\ref{gx}) 
and (\ref{gy}) dominate at long distance. As expected, if the dipole moments of the interband transition are aligned along the axis 
joining the two QDs ($x$-polarization), then the long-range transfer is mainly governed by longitudinal TM-modes, whereas for dipole 
moments orthogonal to this axis ($y$-polarization), it is the transverse TE-modes that dominate in the transfer mechanism. The Gauss 
function in (\ref{G12}) extends to momenta much larger than $k_0$, as $\sigma\ll L_c$. Hence, computing the full mode structure of a 
DBR-MC is important in order to have a quantitatively good account of the transfer mechanism.
In Fig. \ref{fig2} we display the absolute value of $g_k^T(\omega)$ and $g_k^L(\omega)$. Both quantities are characterized by 
one main cavity mode with a cutoff energy $\hbar\omega_0\simeq1.36$ eV. At larger $k$, modulations due to the leaky modes of the 
DBRs are clearly visible. A main difference appear at large momenta, where the T-modes of Fig. \ref{fig2}(a) display 
another strong resonance corresponding to a guided mode close to $k_0$, which is instead absent for L-modes. For $k>k_0$, 
the DBR reflectivity rapidly approaches unity and the Green's functions approach those of a homogeneous medium.

\begin{figure}[ht]
  \centerline{\includegraphics*[width=0.9\linewidth]{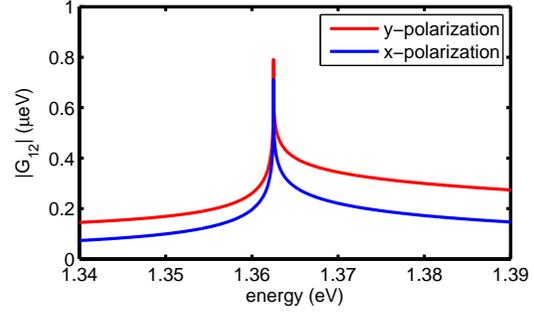}}
  \caption{Transfer amplitudes $|[\hat{G}_{12}(\omega)]_{xx}|$ (blue) and $|[\hat{G}_{12}(\omega)]_{yy}|$ (red) vs energy $\hbar\omega$, at $d=300$ nm.}
  \label{fig3}
\end{figure}

We turn to study the frequency dependence of the transfer amplitudes $[\hat{G}_{12}(\omega)]_{jj}$ ($j=x,y$), which are plotted in Fig. \ref{fig3} for a distance $d=300$ nm. They are strongly resonant with the main cavity mode at $k=0$, indicating that the MC resonance still dominates in the radiative transfer process. The pronounced asymmetry, with significantly larger values at positive detuning, is due to the onset of the radiative transfer through the main cavity mode. Due to the transverse guided mode, $[\hat{G}_{12}]_{yy}$ is about twice as large as $[\hat{G}_{12}]_{xx}$ when off resonance. At negative detuning the transfer is inhibited, although only by one decade because, as seen in Fig. \ref{fig2}, the contribution of leaky and guided modes is practically frequency-independent. This is very important in the light of proposals to use MC polaritons as virtual (off-shell) intermediate states for spin-spin transfer \cite{Quinteiro2006}. We show that resonant (on-shell) transfer may equally take place, independently of QD-MC detuning.

\begin{figure}[ht]
  \centerline{\includegraphics*[width=0.9\linewidth]{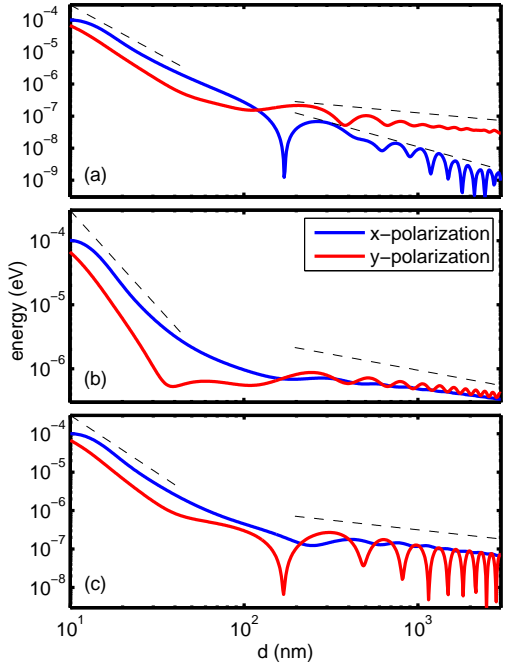}}
  \caption{Transfer amplitudes $|[\hat{G}_{12}(\omega)]_{xx}|$ (blue) and $|[\hat{G}_{12}(\omega)]_{yy}|$ (red) vs QD distance $d$ for (a): $\hbar\omega=\hbar\omega_0-0.03$ eV; (b): $\hbar\omega=\hbar\omega_0$; (c): $\hbar\omega=\hbar\omega_0+0.03$ eV. The dashed lines are guides to the eye for the $d^{-3/2}$ dependence in (a) and for the $d^{-3}$ and $d^{-1/2}$ dependence in the three panels (see text).}
  \label{fig4}
\end{figure}

The transfer amplitudes vs $d$ are plotted in Fig. \ref{fig4} for three values of $\hbar\omega$. 
As a general trend, they show an oscillatory behaviour, as the transfer occurs mainly via a few well-defined momenta given either by the main MC mode or by the guided mode. At very short distance all amplitudes display a $d^{-3}$-dependence. Our full Maxwell formalism thus correctly reproduces the F\"orster transfer mechanism, caused by the electrostatic dipole-dipole interaction \cite{Govorov2005}. Below the cavity resonance $\hbar\omega_0$, the amplitude for $x$-polarized QDs decays as $d^{-3/2}$ whereas the one for $y$-polarized QDs displays an ultralong spatial range characterized by a $d^{-1/2}$-decay. According to Fig. \ref{fig2}, in fact, the main contribution for $x$-polarization comes from the transverse guided mode in $g^t_k$, which in Eq. (\ref{gx}) has an asymptotic $d^{-3/2}$ law. In Eq. (\ref{gy}) instead, $g^t_k$ multiplies the dominant long-range contribution $J_2(kd)$ which decays as $d^{-1/2}$. The largest radiative transfer effect is obtained for $\omega=\omega_0$, namely at resonance with the cavity mode. In this case, the resonant cavity mode is effective for both polarizations and we find a common $d^{-1/2}$ decay. Quantitatively, the rate is about $1~\mbox{ns}^{-1}$ at $d=300$ nm, of the same order as the QD spontaneous radiative rate. The same qualitative trend is displayed when $\omega>\omega_0$, but the overall transfer amplitudes are smaller, as expected also from the results of Fig. \ref{fig3}. Physically, the $d^{-1/2}$ dependence is a direct result of cylindrical wave propagation in 2-D, and should be compared to the corresponding $d^{-1}$ dependence for a bulk medium \cite{Parascandolo2005}.
%, obtained in a bulk medium \cite{Parascandolo2005}.

\begin{figure}[ht]
  \centerline{\includegraphics*[width=1.15\linewidth]{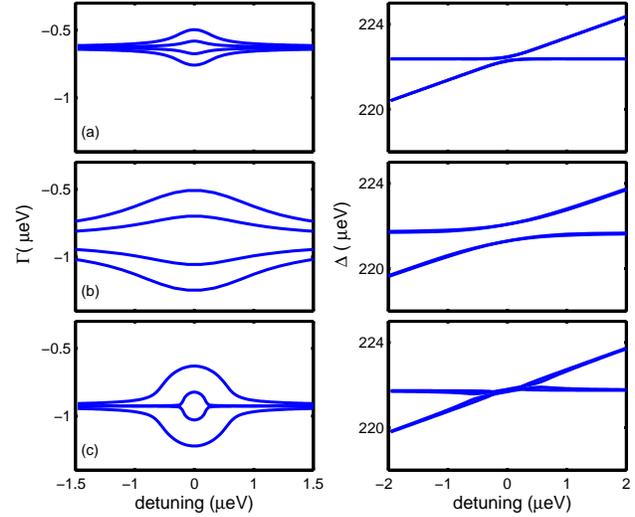}}
  \caption{Imaginary (left) and real (right) part of the energy poles vs QD detuning $\hbar (\omega_1-\omega_2)$ at $d=300$ nm for (a): $\hbar\omega=\hbar\omega_0-0.03$ eV; (b): $\hbar\omega=\hbar\omega_0$; (c): $\hbar\omega=\hbar\omega_0+0.03$ eV.}
  \label{fig5}
\end{figure}

As in the case of QDs in a homogeneous medium \cite{Parascandolo2005,Parascandolo2007}, the effect of radiative transfer is to produce collective many-QD eigenmodes -- the {\em polariton} modes of the system. Their complex energies are the poles of the homogeneous part of Eq. \ref{projected_Dyson} (setting ${\bf E}^0_\alpha=0$). We plot these eigen-energies in Fig. \ref{fig5} for the same three values of the QD-MC detuning as in Fig. \ref{fig4}, as a function of the (much smaller) detuning between the two QDs. As a general trend, the imaginary parts of the eigen energies (corresponding to the radiative rates of the collective modes) take different values at zero QD detuning. One rate is much slower than for the uncoupled QDs, two take intermediate values and one is much faster. This collective effect is enhanced at resonance with the MC, as seen in the comparison of Figs. \ref{fig5} (a)-(c). A similar conclusion is drawn for the real parts. These radiative effects are expected to become significant for a dense assembly of QDs embedded in a planar MC, either randomly or regularly distributed \cite{Ramon2006}, or in the case of few QDs and light confinement in 3-D \cite{Reithmaier2004,Reitzenstein2006,Englund2007,Hughes2007}.

\section{Conclusions}

We have presented a semiclassical theory of radiative excitation transfer between QDs in a planar MC. Due to the 2-D nature of the system, the transfer amplitude decays as $d^{-1/2}$ with the QD distance. The transfer takes place through both the main cavity mode and the guided mode structure of the DBRs, with a rate comparable to the spontaneous emission rate. Additional light confinement, e.g. through photonic crystal waveguides, would enhance the transfer rate, opening the way to applications in quantum information.

\begin{acknowledgement}
We are grateful to Roland Zimmermann and Wolfgang Langbein for enlightening discussions.
\end{acknowledgement}

% \bibliographystyle{pss}
% \bibliography{pss-demo}

\providecommand{\WileyBibTextsc}{}
\let\textsc\WileyBibTextsc
\providecommand{\othercit}{}
\providecommand{\jr}[1]{#1}
\providecommand{\etal}{~et~al.}

\end{document}